\documentstyle[12pt,epsfig]{article}
\newcommand{\beq}{\begin{equation}}
\newcommand{\eeq}{\end{equation}}
\newcommand{\beqar}{\begin{eqnarray}}
\newcommand{\eeqar}{\end{eqnarray}}
\newcommand{\noi}{\noindent}
\begin{document}
\noindent
\hspace*{12.5cm}June. 1997\\
\hspace*{12.5cm}OU-HET 235\\
\hspace*{12.5cm}TOYAMA - 97\\

\vspace*{.2cm}

\begin{center}
  {\LARGE \bf  New Physics and $CP$ Angles Measurement \\
\vspace*{3mm}  at  $B$ Factory}
\end{center}
\vspace*{0.1cm}
\begin{center}
{\large T.~Kurimoto}\footnote{e-mail: krmt@sci.toyama-u.ac.jp}\\
Department of Physics, Faculty of Science,\\
Toyama University,\\
Toyama 930, Japan\\
\vspace*{.2cm}
{\large A.~Tomita}\footnote{present address: Aishin Seiki Jinta-ryo, 
Maebayashi, Toyota, Aichi 473, Japan.}\\
Department of Physics, Faculty of Science,\\
Osaka University,\\
Toyonaka, Osaka 560, Japan\\

\vspace*{1cm}
{\Large\bf Abstract} 
\end{center}
\vspace*{5mm}

We have analyzed how much the $CP$ angles to be measured at B factories can  
deviate from the geometrical ones defined in unitarity triangle 
under the existence of new physics. The measurements are given in rephasing 
invariant form. If KM matrix is not a $3\times 3$ and unitary matrix,  
$\tilde\phi_1$  and $\tilde\phi_3$ is affected, and 
the value of $\tilde\phi_3$ depends on the decay mode.
The deviation is constrained to be less than the experimental precision attained in the 
next decade by the available data of the magnitude of KM matrix elements. Deviation of 
the sum of three angles from $\pi$ cannot be detected unless new physics 
contributes significantly to $b$ decay or $D$ meson system. 
\newpage
\noi
{\large\bf 1. Introduction}
\vspace*{5mm}

In the three generation standard model the $CP$ violation 
phase appears in the quark mixing matrix at the coupling between 
quark charged current and $W$ boson as first pointed out 
by Kobayashi and Maskawa \cite{KM}. The quark mixing matrix, which 
we call KM matrix $V$ hereafter, is a $3\times3$ unitary matrix so that 
the following condition holds :
\beq
{V_{ub}}^* V_{ud} + {V_{cb}}^* V_{cd} + {V_{tb}}^* V_{td} =0.
\eeq  
We have so-called unitarity triangle by expressing the above condition 
in complex plane.
\begin{figure}[hbtp]
  \begin{center}
    \leavevmode
    \epsfig{figure=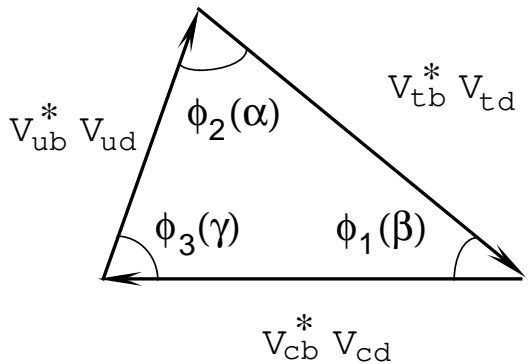, width=7cm}
    \caption{Unitarity triangle}
    \label{fig:unitri}
  \end{center}
\end{figure}
The still unfixed parameters in KM matrix can be determined 
through the measurements of the sides and  the angles of this triangle.
We can test the accuracy of the standard model by over-checking 
the consistency of the triangle. If the test fails, we can explore 
new physics beyond the standard model from the inconsistency \cite{new}. 
This is one of the main aims of the $B$ factory projects which are 
now under way at several experimental facilities, KEK, SLAC and so on.   

New physics can affect the determination of the unitarity triangle through 
(i) $B^0$-$\overline {B^0}$ mixing as in the case of SUSY standard 
models \cite{bbss} , (ii) $b$ decay as in $SU(2)_L \times SU(2)_R \times U(1)$ 
models \cite{lrmd} or (iii) deformation of KM matrix as in 4 generation 
models \cite{fgen} or extra vector-like quark models \cite{vecq}.
It has once said that new physics can be explored by checking 
whether the sum of the three $CP$ angles becomes $\pi$ or not \cite{new}.
But this criterion is not necessarily effective \cite{pith}.
When new physics affects $B^0$-$\overline {B^0}$ mixing alone (case (i)), 
the effects are cancelled in the sum of three angles.  
In the case of (ii) the sum in fact can deviate from $\pi$ \cite{lrmd}. 
The aim of this work is to study the case of case of (iii). It is shown 
under two reasonable assumptions that whether the sum becomes $\pi$ depends 
on what mode is used to measure the angle $\phi_3$ ($\gamma$). The possible 
deviation from $\pi$ is also estimated and found to be 
less than the experimental precision to be attained in the next decade.
Thus the sum of three CP angles will be measured to be  consistent 
with  $\pi$ even within the experimental precision 
even if KM matrix is not $3\times 3$ nor unitary.
The angles measurement alone is not sufficient to explore 
new physics unless new physics significantly affects $b$ decay or 
neutral $D$ meson system. 

Below the effects of new physics on $CP$ angles are discussed  
with the following two assumptions;

\vspace*{3mm}\noi
{\it Assumption I. Quark decay amplitude has the same 
phase as those given by the corresponding tree level $W$ boson exchange diagram 
up to minor corrections except for the $\Delta I=1/2$ penguin type contribution.
\vspace{2mm}

\noi
Assumption II. Both 
$D^0$-$\overline {D^0}$ mixing and $CP$ violation are negligible 
in neutral $D$ meson system.
}

\vspace*{3mm}\noi
The second assumption is necessary since the $CP$ angle $\phi_3$ ($\gamma$) is
measured at $B$ factories by using the decay of $B$ mesons into
a neutral $D$ meson and a $K$ meson.\cite{phit} \
If $D^0$-$\overline {D^0}$ mixing or $CP$ violation is significant in $D$ meson
system, it obscures the determination of $\phi_3$ ($\gamma$).
Let us see if the above assumptions are satisfied taking 
the following models as examples:
\begin{description}
\item [minimal SUSY standard model with no new $CP$ phase]: 
The quark decay is described by the standard $W$ boson processes and 
physical charged Higgs processes. The couplings among quarks and charged 
Higgs are given by KM matrix elements times real factors proportional to quark masses. 
Loop contribution by SUSY particles is suppressed due to SUSY GIM mechanism. 
Higgs loop is suppressed by at least one small Yukawa coupling. 
Therefore both assumptions are satisfied.
\item [multi-Higgs doublets model]:
Both conditions are satisfied as far as flavor changing neutral current 
(FCNC) is suppressed by NFC \cite{nfc} since the couplings among quarks 
and charged 
Higgs are KM matrix elements times real factors proportional to quark masses. 
\item [4 or more generation model]: 
Although the KM matrix is no longer a $3\times 3$ matrix, the assumptions 
are satisfied as far as the 4-th (or higher) generation quarks do not contribute 
significantly to $b \rightarrow s$ penguin process or $D$ meson physics.
\item [vector-like extra quark model]: 
The KM matrix is neither  $3\times 3$ nor unitary. But the assumptions are 
satisfied if FCNC is suppressed.
\end{description}

In the following arguments we examine what are measured as 
the $CP$ angles at $B$ factories under the influence of new physics 
without assuming that KM matrix is $3\times3$ and unitary.  
The measurements as CP angles are given in rephasing invariant form. 
They are compared with the geometrical definitions of the angles in 
unitarity triangle, and we estimate the differences between them.
\vspace*{7mm}

\noi
{\large\bf 2. New physics effects on angles}
\vspace*{5mm}

The angles of the unitarity triangle are geometrically defined as
\begin{eqnarray}
  \label{cpangles}
  \phi_1 \ (\beta) &\equiv &  \mbox{arg}[{V_{cb}}^* V_{cd}]
                                - \mbox{arg}[{V_{tb}}^*V_{td}] - \pi, \\
  \phi_2 \ (\alpha) &\equiv &  \mbox{arg}[{V_{tb}}^* V_{td}]
                                - \mbox{arg}[{V_{ub}}^*V_{ud}] + \pi, \\
  \phi_3 \ (\gamma) &\equiv &  \mbox{arg}[{V_{ub}}^* V_{ud}]
                                - \mbox{arg}[{V_{cb}}^*V_{cd}] + \pi .
\end{eqnarray}
These angles do not necessarily agree with the $CP$ angles to be measured 
in experiments.  
Two of the $CP$ angles corresponding to $\phi_1$ $(\beta)$ and $\phi_2$ 
$(\alpha)$, which we call $\tilde\phi_1$ and $\tilde\phi_2$, respectively,  
are measured through time dependent $CP$ asymmetry of 
the neutral $B$ meson decay into a $CP$ 
eigenstate, $f_{CP}$  \cite{cpang} :
\begin{eqnarray}
 Asy[f_{CP}] &\equiv& 
\frac{\Gamma[B^0 (t) \rightarrow f_{CP}] - 
      \Gamma[\overline{B^0}  (t)\rightarrow f_{CP}]
      }{
      \Gamma[B^0  (t)\rightarrow f_{CP}] + 
      \Gamma[\overline{B^0}  (t)\rightarrow f_{CP}] }\\
 &=&  \frac{2}{(2+c_d)} \left[\mbox{Im}(\frac{q}{p} \rho) \sin (\Delta M_B t)
                              -\frac{c_d}{2}\cos (\Delta M_B t)\right] ,
\end{eqnarray}
where 
\begin{eqnarray}
  \frac{q}{p} \equiv \frac{|M_{12}^B|}{M_{12}^B},&\quad&
  M_{12}^B \equiv \langle B^0|{\cal H}^{\Delta B=2}| \overline {B^0}\rangle, \\
  \rho  \equiv  \frac{A(\overline{B^0}\rightarrow f_{CP})}{
                        A(B^0\rightarrow f_{CP})},&\quad& 
  |\rho|^2 \equiv 1 + c_d, 
\end{eqnarray}
and we have neglected the absorptive part of $\langle B^0|{\cal H}^{\Delta B=2}| 
\overline {B^0}\rangle$, which is a good approximation in $B$ meson system. 
The assumption I given before allows us to express  $\rho$ by KM matrix elements 
up to $\Delta I =1/2$ penguin effect. The rest of the $CP$ angle corresponding to 
$\phi_3$ ($\gamma)$, which we call $\tilde\phi_3$, is measured via direct $CP$ 
violation in $B \rightarrow DK$ decays. Let us see in detail the differences 
between $\phi_i$ and $\tilde\phi_i$ $(i=1,2,3)$. 

The $CP$ angle $\tilde\phi_1$ is measured in $b(\bar b) \rightarrow c\bar c s (\bar s)$ 
decay. A typical hadronic final state is $J/\Psi K_S$. The weak phase of the decay 
amplitude is given by $\mbox{arg}[V_{cb}{V_{cs}}^*]$. There is no direct $CP$ violation 
under the assumption I, i.e. $|\rho|=1$, since no $\Delta I =1/2$ process is involved. 
(There is a $b \rightarrow s $ penguin contribution. But it has almost the same phase 
as that of the tree $W$ boson contribution in the standard model. If 4-th quark or another 
new particle significantly contributes to $b \rightarrow s $ penguin, the assumption I 
can be violated.) 
We have
\beq \left.
\mbox{Im}(\frac{q}{p} \rho)\right|_{J/\Psi K_S} = \mbox{Im}\left[ 
           \frac{|M_{12}^B|}{M_{12}^B} \frac{V_{cb}{V_{cs}}^*}{{V_{cb}}^*V_{cs}}
                 \left(\frac{q_K}{p_K}\right)^* 
        \right],
\eeq
where 
\beq
\frac{q_K}{p_K} = \frac{[(M_{12}^K -(i/2)\Gamma_{12}^K)
                            (M_{12}^{K*} -(i/2)\Gamma_{12}^{K*})]^{1/2}
                            }{M_{12}^K -(i/2)\Gamma_{12}^K},
\eeq
with $M_{12}^K -(i/2)\Gamma_{12}^K \equiv \langle K^0|{\cal H}^{\Delta S=2}|
 \overline {K^0}\rangle$. It is experimentally known that $CP$ violation in 
$K$ meson system is tiny, $O(10^{-3})$, so that we neglect it here. 
Then we can take $ M_{12}^K/\Gamma_{12}^K$ to be real, and 
\beq
\frac{q_K}{p_K} = \frac{|\Gamma_{12}^K|}{\Gamma_{12}^K}
               =  \frac{V_{ud}{V_{us}}^*}{{V_{ud}}^*V_{us}}. \label{pqk}
\eeq
The phase of $\Gamma_{12}^K$ is calculated from  $W$ boson exchange 
tree decay diagram since we neglected $CP$ violation in $K$ meson system. 
Let us define the phase discrepancy between the KM factors as  
\beq
\delta_1 \equiv \mbox{arg}[V_{ud}{V_{us}}^*] -\mbox{arg}[V_{cd}{V_{cs}}^*] + \pi ,
\eeq
where $\delta_1=O(10^{-3})$ in the standard model. 
The phase of ${q_K}/{p_K}$ is often taken to be 
${V_{cd}{V_{cs}}^*}/{{V_{cd}}^*V_{cs}}$ in the literatures by assuming the 
dominance of charm quark contribution in the box diagram of 
of $K^0$-$\overline {K^0}$ mixing as in the standard model. But $\delta_1$ may not 
be negligible if KM matrix is not $3\times 3$ and unitary. There is a possibility that  
the sum of charm quark contribution and new physics contributions 
cancel each other or happen to be almost the same phase with 
that of up quark contribution giving tiny $CP$ violation in $K$ system.  
So it is more appropriate to take ${q_K}/{p_K}$ as in eq.(\ref{pqk}).

We have
\beq
 \frac{Asy[J/\Psi K_s]}{\sin (\Delta M_B t)} = 
 \mbox{Im}\left[  \frac{|M_{12}^B|}{M_{12}^B}
       \frac{V_{cb}{V_{cs}}^*}{{V_{cb}}^*V_{cs}}
       \frac{{V_{cd}}^* V_{cs}}{V_{cd}{V_{cs}}^*} e^{-2i\delta_1}\right]
        = -\sin (\phi_M + 2 \phi_c + 2 \delta_1), \label{phi1}
\eeq
where $\phi_M\equiv \mbox{arg}[M_{12}^B]$, $\phi_c \equiv \mbox{arg}[{V_{cb}}^* V_{cd}]$.
In the case of the standard model $\phi_M = - 2  \mbox{arg}[{V_{tb}}^* V_{td}]$, 
so that the righthand-side of eq.(\ref{phi1}) becomes $-\sin 2 \phi_1$ up to tiny 
correction of $\delta_1$. If a new physics affects $\phi_M$ or $\delta_1$, then 
the $CP$ angle to measure, $\tilde \phi_1 =\phi_M/2 +\phi_c +\delta_1-\pi$ (mod $\pi$), 
can deviate from the angle of the unitarity triangle, $\phi_1$. 

The $CP$ angle $\tilde\phi_2$ is measured in $b(\bar b) \rightarrow u\bar u d (\bar d)$ 
decay. A typical hadronic final state is $\pi\pi$. There is a $\Delta I =1/2$ penguin 
contribution in this decay mode. But it can be removed by isospin 
analysis \cite{iso}. The weak phase of the resulting decay amplitude is 
controlled by KM matrix elements following the assumption I. 
The $CP$ asymmetry is given by 
\beq 
\left.
\mbox{Im}(\frac{q}{p} \rho)\right|_{\pi\pi} = -\mbox{Im}\left[ 
           \frac{|M_{12}^B|}{M_{12}^B} \frac{V_{ub}{V_{ud}}^*}{{V_{ub}}^*V_{ud}}
        \right] = \sin (\phi_M +2\phi_u),
\eeq
where  $\phi_u \equiv \mbox{arg}[{V_{ub}}^* V_{ud}]$.
In the case of the standard model the righthand-side of eq.(11) becomes 
$\sin [2(\pi- \phi_2)]=- \sin 2\phi_2$. With new physics effects on  $M_{12}^B$, 
the $CP$ angle to measure, $\tilde\phi_2 = -\phi_M/2 -\phi_u +\pi$ (mod $\pi$), 
can deviate from $\phi_2$.

The rest of the $CP$ angles $\tilde\phi_3$ is obtained from the decays 
$B \rightarrow DK$,
where $D$ is a  neutral $D$ meson \cite{phit}. The involving quark 
processes are $b \rightarrow c \bar u s, u\bar c s$ and their $CP$ conjugates. 
No penguin process is involved here, so that we can write down the amplitudes 
following the assumption I:
\beq
 A(B^+ \rightarrow D^0K^+) \propto {V_{ub}}^* V_{cs}, \ \ 
 A(B^+ \rightarrow \overline{D^0}K^+) \propto {V_{cb}}^* V_{us}.
\eeq
The neutral $D$ meson is identified by $CP$ eigenstates ($K_S\pi^0$, $K_S\omega$, 
$K_S\phi$ for $CP$ odd state and $K^+K^-$ for even sate) 
or $CP$ non-eigenstates ($K^\pm\pi^\mp$ and so on) into which both $D^0$ and 
$\overline{D^0}$ can decay. The two amplitudes 
$A(B^+ \rightarrow D^0K^+)A( D^0\rightarrow f)$ and 
$A(B^+ \rightarrow \overline{D^0}K^+)A( \overline{D^0}\rightarrow f)$, where $f$ is 
a common final state of neutral $D$ mesons, interferes giving rise to $CP$ violation.
When $f$ is taken to be $K^+K^-$, $K_S\pi^0$, $K_S\omega$ or $K_S\phi$,   
the interference term depends on 
\beqar
-\arg \left[
    \frac{{V_{cb}}^* V_{us}}{{V_{ub}}^* V_{cs}}
    \frac{V_{cs}{V_{us}}^*}{{V_{cs}}^*V_{us}}
    \right]
&=& -\arg \left[
{(V_{cb}}^*V_{cd})( V_{ub} {V_{ud}}^*)( {V_{us}}^* V_{ud})( {V_{cs}} {V_{cd}}^*)
\right] \nonumber \\
&=& \phi_u - \phi_c -\delta_1 +\pi. 
\eeqar
which becomes the CP angle $\tilde\phi_3$ to measure. While if $f$ is taken 
to be $CP$ non-eigenstates, we get  $\tilde\phi_3$ as
\beqar
-\arg \left[
    \frac{{V_{cb}}^* V_{us}}{{V_{ub}}^* V_{cs}}
    \frac{V_{cd}{V_{us}}^*}{{V_{cs}}^*V_{ud}}
    \right]
&=& -\arg \left[
{(V_{cb}}^*V_{cd})( V_{ub} {V_{ud}}^*)
\right] \nonumber \\
&=& \phi_u - \phi_c +\pi. 
\eeqar
If $\pi\pi$ mode is available as $CP$ eigenstate, the measured angle $\tilde\phi_3$ is given as
\beqar
-\arg \left[
    \frac{{V_{cb}}^* V_{us}}{{V_{ub}}^* V_{cs}}
    \frac{V_{cd}{V_{ud}}^*}{{V_{cd}}^*V_{ud}}
    \right]
&=& -\arg \left[
{(V_{cb}}^*V_{cd})( V_{ub} {V_{ud}}^*)( V_{us} {V_{ud}}^*)( {V_{cs}}^* {V_{cd}})
\right] \nonumber \\
&=& \phi_c - \phi_u +\delta_1 +\pi .
\eeqar
There is a difference among the measured angle $\tilde\phi_3$ depending on which of 
the modes is used to identify neutral $D$ meson. 
If $\delta_1$ is larger than the experimental precision, the difference can be seen. 
But the limit on $\delta_1$ is severe as will be shown in the next section,  
so that we need high precision (at least below 10$^\circ$) in  $\tilde\phi_3$ 
measurement to observe the difference. 

Now we have the formulae of the three $CP$ angles to measure:
\begin{eqnarray} \label{phi}
  \tilde\phi_1 &=& \phi_M/2 +\phi_c +\delta_1 ,\\ \nonumber
  \tilde\phi_2 &=& -\phi_M/2 -\phi_u ,\\ \nonumber
  \tilde\phi_3 &=& \pi +\phi_u -\phi_c (\pm\delta_1). \label{angs}
\end{eqnarray}
The sum of three angles becomes $\pi$ up to the correction $\delta_1$. The magnitudes 
of $\delta_1$  is negligible in the three generation standard model but might not be 
neglected when KM matrix is not $3\times 3$ and unitary. The limit on $\delta_1$ 
is discussed in the next section.
The phase of $M_{12}$ cancels between $\tilde\phi_1$ and $\tilde\phi_2$, 
so that the sum does not change if new physics 
affects $B^0$-$\overline {B^0}$ mixing. 

\vspace*{7mm}

\noi
{\large\bf 3. Constraints on  $\delta_{1}$}
\vspace*{5mm}

Here we estimate how much the $\delta_{1}$ can be if KM matrix is not 
in the standard model form, i.e. $3\times 3$ unitary.  
Let us investigate the constraints on $\delta_{1}$ from the present 
experimental values of KM matrix elements 
without assuming $V$ is $3\times 3$ and unitary. 
With usual three generations of quarks and possible 
extra quarks the coupling among quark mass eigenstates and $W$ boson is given as 
\beq
{\cal L}_W = \frac{g_2}{\sqrt{2}}
         \left(\bar u_L\ \bar c_L\ \bar t_L \cdots \right)
         \gamma^\mu V  \left(
         \begin{array}[c]{c}
         d_L \\ s_L \\ b_L \\ \vdots
         \end{array} \right)
        W^+_\mu +\mbox{(h.c.)}.
\eeq
Here $V$ is not necessary unitary nor square, but it can be shown that it is 
at least a part of a larger unitary matrix \cite{bvk}.
Let us suppose there are $N_u$ $u$-type quark mass eigenstates, 
$U_L \equiv (u_L,c_L,t_L,\cdots,\psi_1, \psi_2,\cdots )^T$, which  are 
related to weak eigenstates of $n_f$ ordinary $SU(2)_L$ doublet quarks, 
$U_L^0 \equiv (u_L^0,c_L^0,t_L^0,\cdots )^T$ and $k_u$ extra singlet 
quarks with charge $2/3$, $ \Psi^0 \equiv (\psi_1^0,\psi_2^0,\cdots )^T$ as follows;
\beq
 U_L = \left [ X_u | X_\psi^u \right ]
\left(  \begin{array}[c]{c}
        U_L^0 \\ \hline \Psi^0
  \end{array}\right), 
\eeq
where $N_u \times n_f$ matrix $X_u$ and $N_u\times k_u$ matrix $X_\psi$ 
are composing the unitary matrix. Similar relations hold in $d$-type quarks.
The quark-$W$ coupling becomes as follows;
\begin{eqnarray}
 {\cal L}_W &=& (g_2/\sqrt{2}) \overline {U_L^0} \gamma^\mu D_L^0 W_\mu^+ 
    +\mbox{(h.c.)} \\ \nonumber
      &=&  (g_2/\sqrt{2}) \overline {U_L} \gamma^\mu 
                 (X_u X_d^{\dagger})D_L  W_\mu^+ +\mbox{(h.c.)}.
\end{eqnarray}
The KM matrix $V$ is given by $(X_u X_d^{\dagger})$  which is a $N_u\times N_d$ matrix. 
This KM matrix is not necessarily unitary but is a part of a lager unitary matrix;
\beq
 \Omega \equiv 
\left(
   \begin{array}[c]{cc}
     (X_u X_d^{\dagger})  &  X_\psi^u \\
      {X_\psi^d}^{\dagger} & 0
   \end{array} \right), 
\eeq
which can be shown by simply checking $\Omega^{\dagger}\Omega =1$.

The KM matrix elements $V_{ij}$ ($i=1,2$, $j=1 \sim 3$) are measured through 
semi-leptonic processes of hadrons. We assume that new physics effect is 
negligible in semi-leptonic processes in comparison with the standard model 
contributions. We can adopt the experimental values of the KM matrix elements as 
the genuine ones with this assumption.
This assumption is thought to be reasonable by the following reasons. 
Possible new physics effects on semi-leptonic processes is 
scalar boson exchange  or another gauge boson exchange at the tree level. 
Physical charged Higgs scalars  
can contribute to semi-leptonic processes in multi-Higgs model. But the couplings 
among leptons and Higgs are suppressed due to the small masses of leptons.
(No $\tau$ lepton process is used in the determination of KM matrix elements.) 
So the contribution is negligible even if physical charged scalar is lighter 
than $W$ boson considering its lower bound of mass \cite{PDG}. 
$W_R$ boson has a gauge coupling to leptons 
in $SU(2)_L \times SU(2)_R \times U(1)$ models \cite{lrmd}. 
But the $W_R$ couples to $\nu_R$ which should be much heavier than $b$ quarks 
for the see-saw mechanism to work \cite{ss}. Then $W_R$ cannot contribute 
to the semi-leptonic processes in determining KM matrix elements. 
Loop effects of new physics is thought to be suppressed as most of the new 
particles will be heavier than $W$ boson.  

The present data on KM matrix elements are given as follows \cite{PDG};
\begin{eqnarray*}
     |V_{ud}| &=& 0.9736 \pm 0.0010, \ \  |V_{us}| = 0.2205 \pm 0.0018, \\ 
     |V_{cd}| &=& 0.224 \pm 0.016, \ \  |V_{cs}| = 1.01 \pm 0.18.
\end{eqnarray*} 
Since $V$ is at least a part of unitary matrix $\Omega$, 
we have the following inequalities;
\begin{eqnarray}
  \sum_{i=3}|\Omega_{ui}|^2 & =&
               1-|V_{ud}|^2 -|V_{us}|^2 < 0.0056,\\
  \sum_{i=3}|\Omega_{ci}|^2 &=&
               1-|V_{cd}|^2 -|V_{cs}|^2 < 0.293, \\
  \sum_{i=3}|\Omega_{id}|^2 &=& 
               1-|V_{ud}|^2 -|V_{cd}|^2 < 0.0094,       \\
  \sum_{i=3}|\Omega_{is}|^2 &=&
               1-|V_{us}|^2 -|V_{cs}|^2 <  0.295.
\end{eqnarray}
With these data the following upper bounds are obtained by using Schwarz's inequality;
\begin{eqnarray}
  |\sum_{i=3} \Omega_{ui}{\Omega_{ci}}^*| &\leq& \sum_{i=3} |\Omega_{ui}{\Omega_{ci}}^*|
       \leq \sqrt{ \sum_{i=3}|\Omega_{ui}|^2 \sum_{j=3}|\Omega_{cj}|^2 }
       < 0.040, \\
  |\sum_{i=3} \Omega_{id}{\Omega_{is}}^*| &\leq& \sum_{i=3} |\Omega_{id}{\Omega_{is}}^*|
       \leq \sqrt{ \sum_{i=3}|\Omega_{id}|^2 \sum_{j=3}|\Omega_{js}|^2 }
       < 0.053.
\end{eqnarray}
Here for the discussion below we define 
\beq
\delta_3 = \mbox{arg}[{V_{cs}}^*V_{us}] -\mbox{arg}[{V_{cd}}^*V_{ud}] +\pi.
\eeq
The sum $\delta_1 + \delta_3 $ becomes 0 (mod $2\pi$) as shown  below:
\begin{eqnarray} \label{sum}
  \delta_1 + \delta_3 &=& 
       \mbox{arg}[V_{ud}{V_{us}}^*] -\mbox{arg}[V_{cd}{V_{cs}}^*] + \pi
      +\mbox{arg}[{V_{cs}}^*V_{us}] -\mbox{arg}[{V_{cd}}^*V_{ud}] +\pi \\ \nonumber
            &=& \mbox{arg}[V_{ud}{V_{us}}^*{V_{cs}}^*V_{us}]
               - \mbox{arg}[V_{cd}{V_{cs}}^*{V_{cd}}^*V_{ud}] + 2 \pi \\ \nonumber
           &=& 0  \ \mbox{(mod }2\pi).
\end{eqnarray} 
The unitarity conditions of $\Omega$,
\begin{eqnarray}
    0 &=& V_{ud}{V_{us}}^* +V_{cd}{V_{cs}}^* +\sum_{i=3}\Omega_{id}{\Omega_{is}}^*,\\
    0 &=& V_{ud}{V_{cd}}^* +V_{us}{V_{cs}}^* +\sum_{i=3}\Omega_{ui}{\Omega_{ci}}^*,
\end{eqnarray}
and the magnitudes of $V_{ud}$, $V_{us}$, $V_{cd}$ and $V_{cs}$ give 
\beq
|\delta_1| < 0.25\ (\simeq 14^\circ), \ \ 
|\delta_3| <  0.20\ (\simeq 12^\circ). 
\eeq
The bound on $\delta_3$ 
should be applied also on  $\delta_1$ because of the the identity (\ref{sum}).
It can be found from the above arguments and eq(\ref{angs}) that the 
deviation of the measured angle $\tilde\phi_3$ from the geometrical one  
is less than the experimental precision ($\pm 15^\circ$) by 
the simulation \cite{KEKB} as far as the two assumptions hold. 

\vspace*{7mm}
\noi
{\large\bf 4. Summary}
\vspace*{5mm}

We have analyzed how the $CP$ angles ($\tilde\phi_{1\sim 3}$) to be measured at 
B factories will deviate 
from the geometrical ones ($\phi_{1\sim 3}$) defined in unitarity triangle 
under the existence of new physics.
The new physics effect on $B^0$-$\overline {B^0}$ mixing can be significant 
on $\tilde \phi_1$ and $\tilde \phi_2$ , but is cancelled in the sum of 
these two angles. If KM matrix from is not a $3\times 3$ and unitary matrix,  
$\tilde\phi_1$ and $\tilde\phi_3$ is affected, and the value of $\tilde\phi_3$ 
depends on the decay mode used in the measurement. The deviation 
is constrained to be less than the experimental precision to be attained in the next 
decade by the available data of the magnitude of KM matrix elements. The criterion 
of new physics search to check the sum of the angles is not effective 
unless a new physics significantly affects $b$ decay itself at the tree level 
or through penguin process. If there exists significant contribution of a new physics 
to $D$ meson system, it can affects the determination of $\tilde\phi_3$ with  
$B \rightarrow DK$ process. The obtained value can differ from the 
$\tilde\phi_3$ got by another method like $B_s \rightarrow \rho K_S$ 
where no $D$ meson is involved. It is desirable to get fine  (below 10\%) 
precision and multiple ways of $CP$ angles measurements 
for finding a signature of new physics or constraining one.

\vspace*{3cm}
\begin{center}
  \Large{Acknowledgement}
\end{center}

\noindent
T.K.'s work is supported in part 
by Grant-in Aids for Scientific Research from the Ministry 
of Education, Science and Culture (No. 07804016).

\end{document}